\documentclass[fleqn,twoside]{article}
\usepackage{espcrc2}
\usepackage{times}
\usepackage{mathptm}

\usepackage{graphicx}
\usepackage[figuresright]{rotating}
 \newcommand\beq{\begin{equation}}
 
 \newcommand\eeq{\end{equation}}
 \newcommand\beqn{\begin{eqnarray}}
 \newcommand\eeqn{\end{eqnarray}}
  \def\tspNtot{\tspN_{tot}}
 \def\spN{\sigma^{\pi N}}
 \def\spNtot{\spN_{tot}}
  \def\BpN{B_{\pi N}}
\def\tspN{\tilde\sigma^{\pi N}}

\def\tspN{\tilde\sigma^{\pi N}}
\def\tspNtot{\tspN_{tot}}
\def\tspNel{\tspN_{el}}
\def\tspNin{\tspN_{in}}

\def\GeV{\,\mbox{GeV}}

\def\lsim{\mathrel{\rlap{\lower4pt\hbox{\hskip1pt$\sim$}}
    \raise1pt\hbox{$<$}}}         
\def\gsim{\mathrel{\rlap{\lower4pt\hbox{\hskip1pt$\sim$}}
    \raise1pt\hbox{$>$}}}         

\def\GeV{\,\mbox{GeV}}
\def\MeV{\,\mbox{MeV}}

\title{PCAC and shadowing of low energy neutrinos}

\author{B.Z.~Kopeliovich\address[mpi]{Max-Planck-Institut f\"ur 
Kernphysik,
Postfach 103980, 69029 Heidelberg, Germany}\address[ce]{Departamento 
de Fisica, Universidad Tecnica 
Federico Santa Maria, Valparaiso, Chile}
\address[jinr]{Joint Institute for Nuclear Research, Dubna, Moscow Region 
141980, Russia}
}
       
\begin{document}

\begin{abstract}
 The Adler relation between reactions initiated by neutrinos and pions is
easy to misinterpret as a manifestation of the pion pole dominance. An
axial current, however, cannot fluctuate into a pion, but only to heavy
axial-vector states, since the lepton current is transverse. This is the
miracle of the PCAC hypothesis which dictates a specific conspiracy
between the heavy fluctuations, so that all together they mock the pion
pole. Indeed, the observed $Q^2$ dependence of the axial form factor is
controlled by the effective mass $m_A\sim 1\GeV$, rather than the pion
mass. On the contrary, the onset of nuclear shadowing is governed by the
small pion mass, rather than by the large axial mass scale. This is in
variance with the conventional wisdom which equates the fluctuation
lifetime and the coherence time. For the case of axial current they are
different by almost two orders of magnitude.  As a result, neutrino
interactions are shadowed at very low energies of few hundred MeV, while
energy of about 10~GeV is needed to access nuclear shadowing for the
vector current. On the contrary to naive expectations, nuclear absorption
enhances, rather than suppresses the cross section of coherent
neutrino-production of pions which is the strongest channel (half of the
total cross section) in the black disc limit.
 
\vspace{1pc}
\end{abstract}

\maketitle

\section{Nuclear shadowing of the axial current}

It is known that even weakly interacting particles are shadowed by nuclei,
i.e. the interaction cross section per a bound nucleon is less than on a
free one. Shadowing results from competition of different bound nucleons
in taking part in the interaction. This may produce a sizeable effect only
provided that the interaction cross section is sufficiently large.
Apparently, this is not the case for electromagnetic or weak interactions,
and one may wonder how shadowing happens. The answer has been known for
decades, namely, a weakly interacting particle may develop a hadronic
strongly interacting fluctuation. Of course the probability of such a
fluctuation is tiny, but if its lifetime is longer than the time of
propagation through the nucleus, then once produced this fluctuation
experiences nuclear shadowing which is as strong as in hadronic
interactions.

 The fluctuation lifetime is controlled by the uncertainty principle and
Lorentz time dilation,
 \beq
t_{fluct}=\frac{2\,E}{M_{eff}^2}\ ,
\label{10}
 \eeq
 where $E$ and $M_{eff}$ are the energy and effective mass of the
fluctuation. This is also frequently called coherence time, $t_c$, as it
defines the maximal time interval between two interactions with amplitudes
which have a small phase shift, i.e. are coherent. We will show, however,
that this usual equivalence of the two time scales is not correct for the
axial current. Namely, the coherence time in this case is almost two
orders of magnitude longer than the one given by Eq.~(\ref{10}).

\subsection{PCAC and hadronic properties of neutrino}

The electric charge is not renormalized by the strong interactions due to
vector current conservation, $q_\mu\,V_\mu=0$. One may think it is almost
trivial, since $q_\mu\,\bar{p}\,\gamma_\mu\,n = m_n-m_p\approx 0$.

Data show that the axial charge is also hardly changed by the strong
interactions, pointing at a partial conservation of axial current (PCAC).
This is a very nontrivial phenomenon, since
$q_\mu\,\bar{p}\,\gamma_5\gamma_\mu\,n = m_n+m_p$ is a big quantity. It
can be, however, compensated by the effective pseudo-scalar term,
$g_p\,q_\mu\,\bar{p}\,\gamma_5\,n$, generated in the axial current by the
strong interactions. Such a compensation is possible only if the
pseudo-scalar coupling has a pole, $g_p(q^2)\propto 1/q^2$, corresponding
to a massless pseudo-scalar, Goldstone meson, which must exist to provide
PCAC.  This leads to the Goldberger-Treiman relation between the couplings
$g_{NN\pi}$ and $f_\pi$ and to the PCAC relation for the axial current,
 \beq
\partial_\mu\,A_\mu = f_\pi\,m_\pi^2\,\phi_\pi\ .
\label{20}
 \eeq

The PCAC leads to a relation between the cross sections of interaction of 
neutrino and pion, named after Stephen Adler \cite{adler},
 \beqn
&&\left.\frac{d\sigma(\nu T\to l F)}
{dQ^2\,d\nu}\right|_{Q^2=0} 
\nonumber\\ &=& 
\frac{G^2}{2\pi^2}\,f_\pi^2\,
\left({1\over\nu}-{1\over E}\right)\,
\sigma(\pi T\to F)\ .
\label{30}
 \eeqn
  Here $F$ is the hadronic finals state produced on target $T$ either by
the neutrino (left-hand side), or by the pion (right-hand side); $E$ is
the energy of the neutrino; $\nu$ is the transferred energy.

The structure of this relation is very similar to one suggested for the
vector current by the vector dominance model (VDM),
 \beqn
&&\left.\frac{d\sigma(\nu T\to l F)}
{dQ^2\,d\nu}\right|_{Q^2=0} =
\frac{G^2}{4\pi^2}\,f_\rho^2\,
\frac{|\vec q|}{E_\nu^2}\,
\frac{Q^2}{(Q^2+m_\rho^2)^2}
\nonumber\\ &\times& 
\frac{1}{1-\epsilon}\,
\Bigl[\sigma_T(\rho T\to F)
+\epsilon\,\sigma_L(\rho T\to F)\Bigr]\ ,
\label{40}
 \eeqn 
 where $\epsilon$ is the $\rho$ polarization which might be transverse (T) 
or longitudinal (L).
 
In both cases of axial, Eq.~(\ref{30}), and vector, Eq.~(\ref{40}),
currents the neutrino cross section is proportional to the hadronic,
$\pi$, or $\rho$, cross sections, which are subject to a substantial
shadowing if the target $T$ is a nucleus. The intuitive light-cone
interpretation of the VDM considers different Fock components of the
vector current assuming that vector mesons, in particular $\rho$, are the
dominant ones. If the lifetime of these fluctuations, Eq.~(\ref{10}) is
sufficiently long, shadowing effects are at work\footnote{Note that in
this case $t_{fluct}=t_c$.}

This is why it is tempting to interpret the Adler relation Eq.~(\ref{30})
as a manifestation of pion dominance, i.e. a fluctuation of the axial
current to a pion which interacts with the target. Such a fluctuation,
however, is forbidden. Indeed, the Lorentz structure of the hadronic
current in this case would be,
 \beq
A_\mu=f_\pi\,
\frac{q_\mu}{Q^2+m_\pi^2}\,
A(\pi T\to F)\ .
\label{50}
 \eeq 
 The factor $q_\mu$ acting on the transverse lepton current gives zero, or 
a tiny lepton mass, $q_\mu\,\bar l\,\gamma_5\gamma_\mu\,\nu = m_l$.

Thus, a neutrino can produce only heavy axial fluctuations, like
$a_1$-meson, $\rho\pi$ pair, etc. This looks like a miracle that all those
states add up and act like a pion, and this fine tuning must be
independent of the target. Nevertheless, this is what the PCAC relation is
about. Unfortunately, the details of this phenomenon are beyond our
knowledge of hadronic dynamics, and we should treat it as a hypothesis
aimed at explanation of the observation of the nearly conserved axial
coupling.

\begin{figure}[tbh]
 \vskip -0.6 cm \hskip -0.0 cm 
\rotatebox{1.5}{
\includegraphics[width=0.46\textwidth]{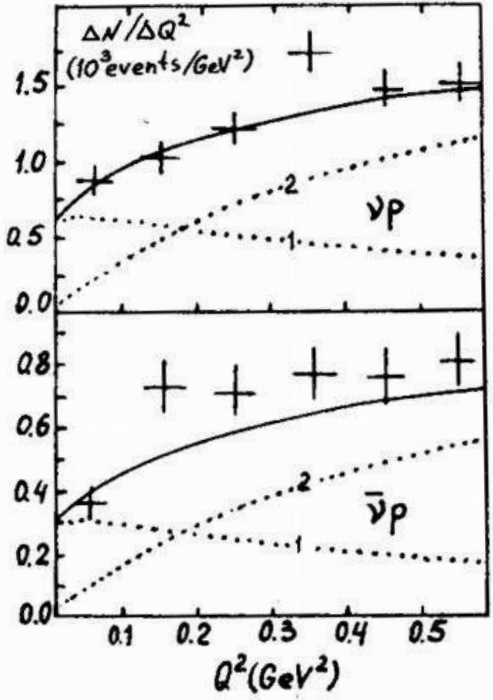}
}
 \vskip -1.0 cm
 \caption{$Q^2$ dependence of the total $\nu p$ and $\bar \nu p$ cross
sections. The dotted curves, {\bf 1} is an extrapolation to nonzero $Q^2$
of the Adler relation; {\bf 2} is calculation employed VDM for the vector
and axial current. Solid curve show the sum of the two contributions. Data
are taken from \cite{km}.}
 \label{q2-dep} 
 \vskip -0.3 cm
 \end{figure}

To be convinced that PCAC is indeed provided by heavy hadronic
fluctuations, rather than a pion, one can look at the $Q^2$-dependence of
the cross section. It is given by the fluctuation propagator, therefore
the width of the $Q^2$ distribution gives the effective fluctuation mass.  
Indeed, data depicted in Fig.~\ref{q2-dep}
 clearly demonstrate the following features:

\begin{itemize}

\item There exists a longitudinal part of the cross section which does not 
vanish at $Q^2\to 0$;

\item Its magnitude well agrees with the Adler relation;

\item
The axial mass parameter controlling the $Q^2$ dependence, is large, 
$m_A^2\sim 1\GeV^2$, two orders of magnitude larger than the pion mass 
squared.

\end{itemize}

Another unusual property of the axial current which is worth mentioning:  
there is no vector dominance in this case, though it is tempting to assume
that like for the vector current the contribution of the lowest axial
vector meson dominates. This assumption has led, however, to a problem
called Piketty-Stodolsky puzzle \cite{ps}. Namely, the cross section of
the off-diagonal diffractive process $\pi N\to a_1 N$ is so small that
provides only a few percent of the observed cross section of
neutrino-production of pions. The main contribution comes from the
$\rho\pi$ cut \cite{bk,km}.

\subsection{Shadowing of low energy neutrinos}

As far as the effective mass of a typical hadronic fluctuation of a
neutrino is as large as $1\GeV$, quite a high energy, $\nu\gsim 10\GeV$,
is needed, to make the fluctuation lifetime Eq.~(\ref{10}) comparable with
the radii of heavy nuclei. Thus, one may jump to a conclusion that there
should be no shadowing at low energies.

However, this conclusion is not correct. It is based on the usual wisdom
that the fluctuation lifetime and the coherence time are equivalent
quantities, what is not true in this case. Indeed, for elastic
neutrino-production of pions, $\nu p \to l\,\pi N$, the longitudinal
momentum transfer, $q_L=(m_\pi^2+Q^2)/2\nu$, is rather small at $Q^2\lsim
m_\pi^2$, i.e. the coherence time, $t_c=1/q_L$ is very long even at low
energy of few hundred $\MeV$. This is actually what matters for shadowing.
As for the fluctuation lifetime, it is indeed much shorter.

This is a result of the nontrivial origin of PCAC. Impossibility to have a
pion in intermediate state leads to the dominance of off-diagonal
processes, like $\nu\to \mu a_1$ and $a_1N\to \pi N$. Same happens for the
vector current, if one considers, for example, $\rho$ photoproduction via
intermediate excitation $\rho'$: $\gamma\to\rho'$ and $\rho' N\to\rho N$.
Such a off-diagonal contribution is negligibly small for the vector
current, but is a dominant one for neutrinos.  Only for diagonal
transitions the fluctuation lifetime and the coherence time are equal,
$t_{fluct}=t_c$. For neutrino interactions the former controls the $Q^2$
dependence of the cross section, while the latter governs shadowing.

\begin{figure}[tbh]
 \vskip -0.7 cm \hskip 0.1 cm 
\includegraphics[width=0.48\textwidth]{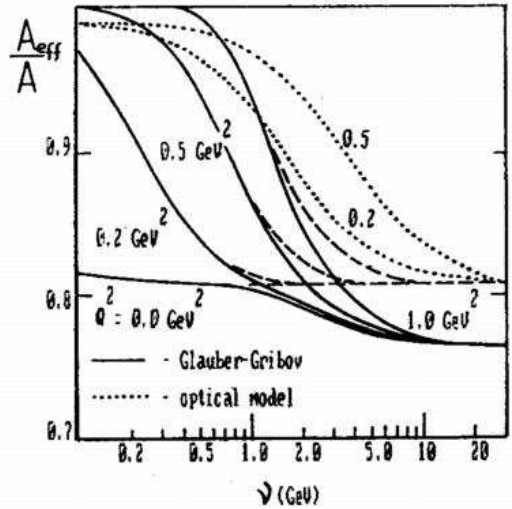}
 \vskip -1.0 cm
 \caption{Energy dependence of the ratio of total neutrino cross sections
on neon and nucleon at different $Q^2$ \cite{k1,k2}. Dashed curves
correspond to the Glauber approximation. Solid curves are corrected for
the Gribov's inelastic shadowing. Dotted curves show the results of the
Bell's optical model \cite{bell}.}
 \label{fig1}
 \end{figure}

The total neutrino-nucleus cross section was calculated in \cite{k1,k2}
taking into account the phase shifts between different points of
interaction (effects of coherence). The calculations are performed using
the Glauber approximation and also including the Gribov's inelastic
corrections (important only at high energies). The results for neon are
depicted in Fig.~\ref{fig1} by solid curves as function of energy for
different $Q^2$. 
 As was anticipated, a rather strong shadowing occurs at small $Q^2$ in
the low energy range of hundreds MeV. This is an outstanding property of
the axial current.

\begin{figure}[tbh] 
\vskip -0.2 cm
\hskip 0.1 cm
\includegraphics[width=0.45\textwidth]{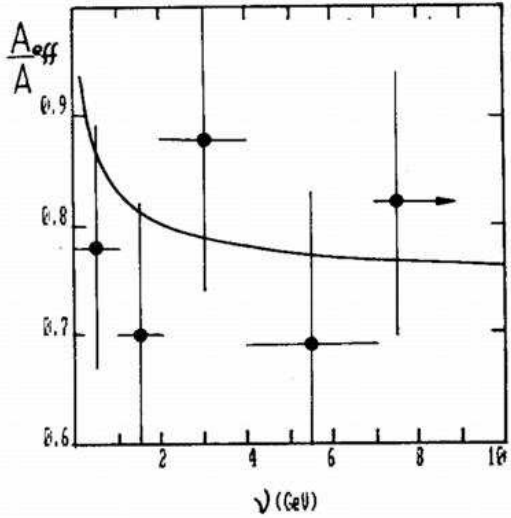} 
\vskip -1.0 cm
 \caption{The neon to proton ratio of the total neutrino cross sections,
calculated in \cite{k1,k2} for $x<0.2$ and $Q^2<0.2\GeV^2$. The data are 
from \cite{wa59}.}
\label{fig2} 
 \end{figure}

The calculations are in a good agreement with available data from the BEBS 
experiment at CERN \cite{wa59} depicted in Fig.~\ref{fig2}. 
 Although with a rather poor statistics the data confirm an early onset of
nuclear shadowing at energies below $1\GeV$.

\section{Diffractive neutrino-production of pions}

\subsection{Pion production on free nucleons} The differential cross
section of this reaction on a nucleon target is given by the Adler
relation,
 \beqn
&&\left.\frac{d\sigma(\nu N\to \mu\pi F)}
{dQ^2\,d\nu\,d^2k_T}\right|_{Q^2=0} \nonumber\\ &=& 
\frac{G^2}{2\pi^2}\,f_\pi^2\,
\left({1\over\nu}-{1\over E}\right)\,
\frac{d\sigma_{el}^{\pi N}}{d^2k_T}\ .
\label{60}
 \eeqn

This expression can be extrapolated to nonzero values of $Q^2$ using
a form factor parametrized usually in a pole form,
 \beq
F_N(Q^2)=\frac{1}{1+Q^2/m_A^2}\,
\label{70}
 \eeq
 which well fits data with $m_A$ close to the mass of $a_1$ meson. This
fact is treated sometimes as an evidence for axial-vector meson dominance,
but as was mentioned, that would lead to the Piketti-Stodolsky puzzle. It
turns out, however, that the $\rho\pi$ cut provides the dominant
contribution to the cross section, leading to a pole-like $Q^2$
dependence. Indeed, the form-factor calculated using the Deck model
\cite{deck} reads \cite{bk},
 \beqn
F_N^{\rho\pi}(Q^2)&=&\frac{(m_\rho^2+m_\pi^2)^2}
{m_\pi^2+Q^2}\,
\ln\left[1+\frac{m_\pi^2+Q^2}
{(m_\rho+m_\pi)^2}\right]\nonumber\\
&\approx&
\frac{1}{1+Q^2/m_A^2}\,
\label{80}
 \eeqn
 where $m_A^2=2(m_\rho^2+m_\pi^2)$ is indeed very close to the $a_1$ mass.
Thus, the cut contribution mocks the $a_1$ pole.

\subsection{Coherent production off nuclei}

It is the commonly accepted terminology to call {\it coherent} a process
which leaves the nuclear target intact in its ground state. The cross
section for coherent production of pions reads \cite{bk,ik},
 \beqn
  && \frac{d\sigma\,(\nu A\to\mu\pi A)}
  {d\nu\,dQ^2\,dk_T^2} \nonumber\\&=&
  \frac{G^2 f_\pi^2}{2\pi^2}\,
  \left({1\over \nu}-{1\over E}\right)\,
  F_N^2(Q^2)\,\Phi_{coh}(k_T,k_L)\ ,
 \label{90}
 \eeqn
 where
 \beqn
&&  \Phi_{coh}(k_T,k_L) = 
  \frac{|\tspNtot|^2}{16\pi}\,
  \left|\int\!\!d^2b\,e^{i\vec b\cdot\vec k_T}
  \int\limits_{-\infty}^\infty 
  dz\,e^{i z k_L}\right.
 \nonumber\\ &\times& \left.
  \rho_A(b,z)\,
  \exp\!\left[-\frac{\tspNtot}{2}\,
 T_z(b,z)\right]\,\right|^2\ .
 \label{100}
 \eeqn
 Here the integration is taken over impact parameter $\vec b$ and
longitudinal coordinate $z$; $\vec k_T$ and $k_L\approx
(Q^2+m_\pi^2)/2\nu$ are the transverse and longitudinal momentum
transfers; $\rho_A$ is the nuclear density and $T_z(b,z)= \int_z^\infty
dz\, \rho_A(b,z)$ is the nuclear thickness; $\tspNtot =
\spNtot\cdot(1-i\alpha_{\pi N})$, where $\alpha_{\pi N}$ is the real to
imaginary part ratio for the forward elastic $\pi N$ amplitude.

\begin{figure}[tbh] 
\vskip -0.2 cm
\hskip -0.1 cm
\includegraphics[width=0.47\textwidth]{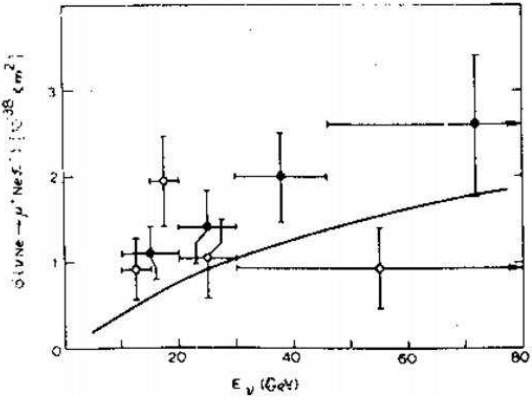} 
\vskip -1.0 cm
\caption{Energy dependence of the cross section of coherent 
neutrino-production of pions on neon. Data are from \cite{data1,data2}.}
\label{total} 
 \end{figure}

The result of calculations \cite{bk} for the cross section Eq.~(\ref{90})  
for neon target is depicted in Fig.~\ref{total} as function of energy.
Comparison with data demonstrates good agreement. 

The phase factor in (\ref{100})  oscillates at low energy and suppresses
the coherent pion production, as one can see in the figure. On the
contrary, at high energies, in the limit of $k_L\,R_A\ll1$ the phase
factor in (\ref{100}) can be neglected and the differential cross section
Eq.~(\ref{90}) of neutrinoproduction of pions is proportional to the cross
section of elastic pion-nucleus collisions, i.e.
Eqs.~(\ref{90})-(\ref{100}) become equivalent to the Adler relation.  For
very heavy nuclei (black disk limit) the elastic $\pi-A$ cross section
reaches the maximum, $\pi\,R_A^2$. Note that the popular Rein-Sehgal model
\cite{rs} predicts zero cross section in this limit contradicting quantum
mechanics and the Adler relation.

\subsection{Incoherent production}

Neutrino can produce diffractively a pion on a bound nucleon, $\nu N\to
\mu\pi N$, and knock the nucleon out of the Fermi surface. In this case 
the bound nucleons act incoherently, and the nucleus breaks up into 
fragments. Such a process has a weaker $A$-dependence ($A^{1/3}$ compared 
to $A^{2/3}$ for coherent production), but is not suppressed by the 
nuclear form factor at low energy, where it turns out to be the dominant 
contribution to pion production.

The cross section of incoherent (quasielastic) neutrinoproduction of pions 
reads \cite{bk,hkn,ik},
 \beqn
 &&\frac{d\sigma\,(\nu A\to\mu\pi A^*)}
 {d\nu\,dQ^2\,dk_T^2}\nonumber\\ &=&
  \frac{G^2 f_\pi^2}{\pi^2}\,
  \frac{EE'-Q^2/4}{2 E^2 |\vec q\,|}\,
  F_N^2(Q^2)\,\Phi_{inc}(k_T,k_L)\,,
 \eeqn
 where $E'=E-\nu$, and
 \beqn
  &&\Phi_{inc}(k_T,k_L) =
  \frac{|\tspNtot|^2}{16\pi}\,
  \exp{\left(-\BpN\,k_T^2\right)}
\nonumber\\ &\times&
\int\!\!d^2b\,
\left\{\frac{1}{\tspNin} 
  \left[1-\exp\!\left(-\tspNin\,T_A(b)\right)\right]\right.
\nonumber\\ &+&
    \frac{\tspNtot\left(\tspNin-\tspNel\right)}{2\tspNel}\,
    \int\limits_{-\infty}^\infty\!
   dz_1\,\rho_A(b,z_1)\,e^{-i\,k_L\,z_1}
   \nonumber\\ &\times&
    \exp\left[
     -\frac12\,\tspNtot\,T_z(b,z_1)\right]
    \int\limits_{z_1}^{\infty}\!\!dz_2\,
  \rho_A(b,z_2)\,e^{i\,k_L\,z_2}
  \nonumber\\ &\times&
  \left. 
    \vphantom{\inf}
    \exp\left[-\frac12\,
   \left(\tspNin-\tspNel\right)\,T_z(b,z_2)\right]
    \right\}\,.
 \eeqn
 There are two terms in this expression, the fist one corresponds to the 
low energy limit when the oscillations terminate the second term. This 
first term has pure classical form with no interferences. In the high 
energy limit, when $k_L\ll 1/R_A$, the whole expression takes the form of 
the cross section for quasielastic pion-nucleus scattering.

\begin{figure}[tbh] 
\vskip -0.6 cm
\hskip -0.4 cm
\includegraphics[width=0.47\textwidth]{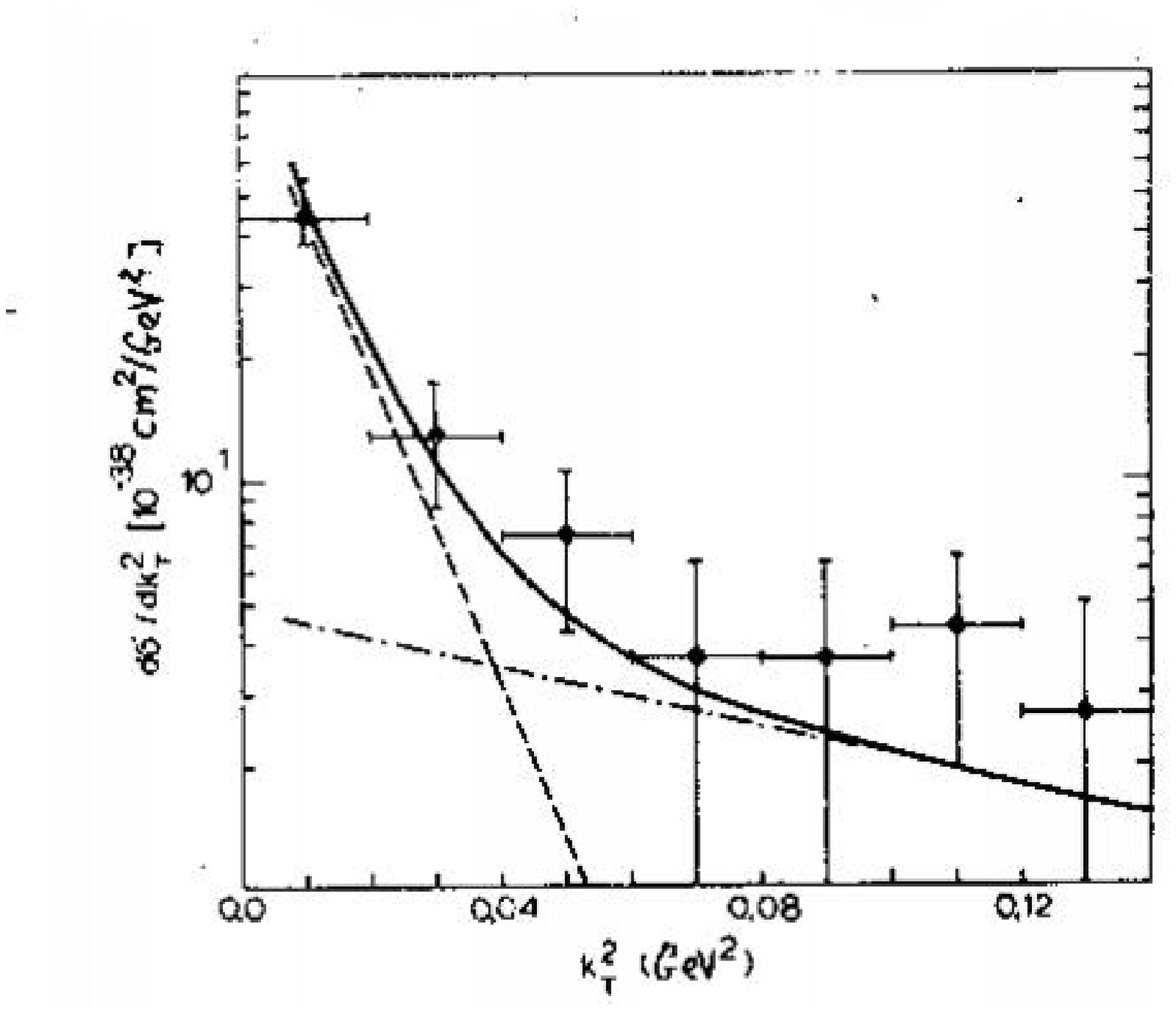} 
\vskip -1.0 cm
 \caption{$k_T^2$ dependence of the cross sections of neutrinoproduction 
of pions on neon, coherent (dashed) and incoherent (dashed-dotted). The 
sum of the cross section is shown by solid curve. The data are from 
\cite{data1}.}
\label{diff} 
 \end{figure}

Example of numerical calculations is presented in Fig.~\ref{diff} where
both coherent and incoherent cross sections are compared with data
\cite{data1} for neon.

\section{Conclusions}

To summarize, we highlight the main observations of this talk.

\begin{itemize}

\item
 PCAC is a hypothesis suggested by the observed small effect of 
renormalization of 
the axial charge. Although this hypothesis has well passed the low energy 
tests, like the Goldberger-Treiman relation, further tests are much 
encouraged. In particular, neutrino interactions providing an intensive 
source of axial current should be used for this purpose \cite{km}.

\item
 One may mistreat PCAC and the Adler relation as a manifestation of pion 
dominance for the axial current. However, neutrino cannot emit a pion 
fluctuation because of transversity of the lepton current.

\item
 The dispersion relation for the axial current is dominated by heavy
states with mass of the order of $1\GeV$. Probably the most nontrivial and
intriguing property of PCAC is that all those heavy states conspire in a
way that they mock the pion pole contribution. The observed $Q^2$
dependence of the cross section indeed confirms that the effective mass is
$m_A\sim 1\GeV$.

\item The axial current exhibits neither pion dominance, nor axial-vector
meson ($a_1$)  dominance. The latter is due to the strong suppression of
diffractive $\pi\to a_1$ transitions observed in data. The important
contribution to the dispersion relation for the axial current is the
$\rho\pi$ cut. The corresponding axial form-factor has a form imitating
the $a_1$ pole contribution.

\item
 In spite of the dominance of heavy fluctuations, the onset of nuclear
shadowing for neutrinos is controlled by the pion mass. This seems to
contradict the intuition based on the conventional wisdom suggesting that
the coherence time and fluctuation lifetime are the same things. However,
in the case of off-diagonal processes, like one under discussion, this is
not true. The lifetime of fluctuations is much shorter than the coherence
time.

\item
 The coherence time for the axial current at small $Q^2$ turns out to be
about two orders of magnitude shorter than for the vector current. This
nontrivial observation leads to an onset of nuclear shadowing for
neutrinos at extremely low energies, hundreds of MeV.

\item
 The strong channel of neutrino interaction, coherent pion production is  
enhances, rather than suppressed by nuclear absorption. The heavier and 
less transparent is the nucleus, the more pions are produced coherently.
In the black disk limit (everything is absorbed) this is the strongest 
channel of neutrino-nucleus interaction.

\item
 Although coherent neutrino-production is the dominant source of pions at
high energies, the incoherent process takes over at low energies, where 
the coherent process is substantially suppressed by the nuclear form 
factor.

\end{itemize}

\end{document}